\documentclass[a4paper,fleqn]{cas-sc}

\makeatletter
\renewenvironment{figure}[1][tbp]{%
  \@float{figure}[#1]%
}{%
  \end@float%
}
\renewenvironment{table}[1][tbp]{%
  \@float{table}[#1]%
}{%
  \end@float%
}
\makeatother

\usepackage[authoryear]{natbib}
\usepackage{booktabs}
\usepackage{array}
\usepackage{multirow}
\usepackage{siunitx}
\usepackage{url}
\usepackage{listings}
\usepackage{xcolor}
\usepackage{upquote}
\usepackage[normalem]{ulem}

\ExplSyntaxOn
\cs_set:Npn \__first_footerline: {}
\ExplSyntaxOff

\begin{document}
\let\WriteBookmarks\relax
\def\floatpagepagefraction{1}
\def\textpagefraction{.001}

\shorttitle{FiLark: streaming-first DAS framework}
\shortauthors{Li et al.}

\title[mode=title]{FiLark: a streaming-first software framework for end-to-end
exploration, annotation, and algorithm integration in distributed acoustic sensing}

\author[1]{Jintao Li}[orcid=0000-0002-1022-9170]
\ead{lijintao@zju.edu.cn}
\credit{Conceptualization, Software, Methodology, Writing - original draft.}

\author[1,2]{Weichang Li}[orcid=0000-0002-8444-9395]
\cormark[1]
\ead{weichangli@zju.edu.cn}
\credit{Supervision, Conceptualization, Algorithm coding, Writing - review \& editing, Funding acquisition.}

\author[2]{Kai Tong}
\ead{tongkai@zju.edu.cn}
\credit{Validation, Investigation, Writing - review \& editing.}

\author[3]{Xiangyu Guo}
\ead{guoxy828@zju.edu.cn}
\credit{Validation, Investigation, Writing - review \& editing.}

\cortext[cor1]{Corresponding author.}

\affiliation[1]{organization={State Key Laboratory of Ocean Sensing \& Ocean College,
  Zhejiang University},
  city={Zhoushan},
  postcode={316021},
  country={China}}

\affiliation[2]{organization={College of Information Science and Electronic Engineering,
  Zhejiang University},
  city={Hangzhou},
  country={China}}

\affiliation[3]{organization={College of Computer Science and Technology,
  Zhejiang University},
  city={Hangzhou},
  country={China}}

\begin{abstract}
Distributed acoustic sensing (DAS) systems generate continuous, ultra-high-channel-count 
data streams at rates that exceed the capabilities of conventional batch-oriented analysis 
frameworks. As a result, essential tasks such as interactive exploration of long-duration 
recordings, scalable event annotation, and real-time algorithm-in-the-loop monitoring remain 
inadequately supported by workflows built around manually selected data segments and offline 
processing. This paper presents FiLark (Fiber Lark), a Python framework that applies a
\emph{streaming-first} principle uniformly across data access, signal
processing, visualization and monitoring for DAS. Instead of operating on manually 
selected data segments, FiLark presents any DAS sources-including continuous multi-file 
recordings-as a unified stream and builds all system components around that abstraction.
An OpenGL-based ring-buffer renderer enables interactive browsing and visualization of 
arbitrarily long recordings with constant memory usage. An integrated annotation interface
supports event labeling directly within continuous data streams, facilitating the creation of 
reproducible machine-learning-ready labeled datasets without offline preprocessing.
The signal processing library includes temporal, spatial, spectral, and
decomposition-based operators, with both CPU implementations and GPU-accelerated variants
via PyTorch, alongside stateful chunked execution that preserves processing continuity and 
application semantics across segment boundaries. A standardized monitor interface further 
integrates streaming detectors and learning-based models into the visualization workflow.
By sharing a common streaming abstraction across all layers, FiLark allows processing
configurations and workflows developed interactively to transfer directly to scalable
production pipelines without modification.
\end{abstract}

\begin{highlights}
\item FiLark applies a streaming-first design principle across its entire
      stack---I/O, signal processing, algorithm integration,
      visualization and monitoring---so that every layer operates on incremental 
      streaming data rather than pre-loaded segments.
\item A ring-buffer OpenGL renderer enables interactive browsing of arbitrarily
      long DAS recordings at real-time frame rates, with a render-time display 
      filter whose latency is independent of dataset size.
\item An annotation system integrated into the streaming viewer allows event
      labeling within continuous recordings without segment pre-selection,
      creating a direct path from data exploration to labeled,
      machine-learning-ready datasets.
\item A comprehensive signal processing library covers the principal DAS
      operator families with CPU, GPU-accelerated (PyTorch), and stateful
      overlap-aware chunked execution variants.
      A standardized monitor interface connects this processing foundation to
      streaming event detection within the same viewing session.
\end{highlights}

\begin{keywords}
Distributed acoustic sensing \sep scientific software \sep streaming data \sep
interactive visualization \sep annotation \sep signal processing \sep Python
\end{keywords}

\maketitle

\section{Introduction}
\label{sec:intro}

Distributed acoustic sensing transforms standard optical fiber into a spatially dense
array of acoustic strain sensors and has emerged as a powerful acquisition
modality in seismology, infrastructure monitoring, and ocean observation
\citep{LindseyMartin2021}.
Modern interrogator units routinely acquire data from thousands of channels at 
sampling rates ranging from tens to thousands of hertz, producing continuous 
channel-time recordings at scales that challenge conventional data-processing 
workflows \citep{sladen2019distributed, lior2021detection}. In typical deployments, 
data volumes can reach tens of gigabytes per hour and increase substantially in 
high-rate acquisition settings. For example, a system recording 6,000 channels 
at 1,500~Hz in single precision produces approximately 3~TB per day, far exceeding
the practical memory capacity of standard workstations. Consequently, long-duration 
deployments such as submarine cable monitoring or persistent infrastructure surveillance
can accumulate tens to hundreds of terabytes over a single acquisition campaign.
Applications such as seismic event detection, traffic monitoring, ocean observation, 
and infrastructure surveillance all depend on identifying transient signals within 
large continuous DAS recordings \citep{prechelt2018das, lellouch2021low, wang2018das_traffic}.
In many operational settings—including earthquake early warning and real-time pipeline 
integrity monitoring—data must be processed continuously as it is acquired to enable 
timely response. Even in offline analysis workflows, the ability to efficiently 
navigate and interrogate multi-day continuous recordings is a fundamental requirement 
for practical data exploration, quality control, and event analysis.

The software ecosystem available to DAS researchers has not evolved at the same pace 
as the scale and continuity of modern DAS acquisitions.
Existing open-source frameworks, including
DASPy \citep{hu2024daspy}, DASCore \citep{chambers2024dascore},
and xdas \citep{trabattoni2025xdas}, provide 
valuable functionality for array manipulation, format conversion, and 
signal processing, but they largely follow a \emph{batch-first} workflow.
In these systems, data are first selected and loaded into memory before processing 
and visualization are performed on the resulting static block. While effective 
for post hoc analysis of short, pre-identified segments, this paradigm becomes 
limiting during the earlier stages of analysis, where researchers must explore 
long continuous recordings to discover, inspect, and annotate candidate events 
for downstream processing. Although xdas includes the companion annotation tool 
Xpick, which offers a web-based interface for seismic phase picking, it is similarly 
designed for short preloaded segments, and its documentation cautions against loading 
more than a few minutes of data at once. As a result, efficient interactive navigation 
of continuous multi-day recordings and integrated annotation within streaming 
workflows remain largely unsupported by existing DAS software tools.

Beyond the challenge of interactive browsing itself, the lack of integrated 
continuous-record navigation has important downstream consequences for data-driven 
DAS analysis. Labeled DAS datasets remain scarce despite the field being a 
prototypical large-scale sensing problem characterized by high acquisition rates, 
geographically diverse environments, and a wide variety of signal classes, 
including seismic arrivals, traffic activity, ship passages, and ocean acoustic 
phenomena. These characteristics make DAS a natural target for machine-learning 
approaches. Across geophysics, data-driven methods have already transformed 
practice: deep neural networks now routinely outperform classical algorithms in 
seismic phase picking, event characterization, and noise classification 
\citep{mousavi2022deeplearning, yang2022sneakyscience}, while large supervised 
models are widely used in exploration geophysics for tasks such as fault 
segmentation, facies classification, seismic enhancing and denoising, and
seismic interpretation at industrial scale \citep{wu2019faultseg, li2026high}.
Crucially, these advances have been enabled by large curated benchmark datasets
containing millions of labeled examples \citep{mousavi2019stead}.
In contrast, publicly available DAS datasets with meaningful signal diversity 
typically contain only a few thousand annotated events. This limitation reflects 
not only the intrinsic complexity of DAS data, but also a fundamental tooling gap. 
Efficient creation of labeled DAS datasets requires seamless navigation of long 
continuous recordings together with integrated annotation workflows. 
When browsing, event extraction, and labeling are separated across incompatible 
tools, the workflow becomes fragile and difficult to reproduce. Exporting image 
snippets to generic annotation software such as LabelMe \citep{russell2008labelme}
often discards the physical and temporal 
metadata needed to map labels back to the original measurements, while ad hoc 
scripting between stages does not scale to large datasets or collaborative workflows.

A third challenge lies in scalable signal processing. Because DAS recordings are 
typically far too large to fit into memory, chunked and overlap-aware execution is 
not merely an optimization but a fundamental architectural requirement. Many commonly 
used operations—including infinite impulse response (IIR) filtering, running normalization, and detectors with 
temporal context extending across chunk boundaries—require explicit propagation of 
processing state between successive segments. In most existing batch-oriented 
frameworks, these boundary conditions and state-management details are left entirely 
to the user, complicating the development of reliable large-scale workflows. 
At the same time, GPU acceleration can reduce execution times by one to two orders 
of magnitude for filter-intensive and transform-heavy processing pipelines, yet 
native GPU support remains largely absent from current open-source DAS software. 
A signal-processing framework designed for DAS research should therefore provide 
first-class support for CPU and GPU execution, stateful streaming operators, and 
correct boundary handling, allowing identical processing chains to operate seamlessly 
in both interactive exploration and large-scale batch execution. A standardized 
interface for integrating custom detectors and monitoring algorithms further lowers 
the barrier to incorporating new methods into the broader processing and visualization 
workflow.

FiLark was developed to address these limitations through a unified streaming-first 
architecture. Every layer of the framework—from data access and signal processing to 
visualization, annotation, and monitoring—is designed to operate on incremental data 
streams rather than preloaded segments. This architectural consistency enables 
exploration, annotation, processing, and detection to share the same abstractions 
and execution pipeline. Processing configurations developed interactively on a small 
region of interest can be applied directly to full-recording batch execution without 
code modification. Detection algorithms registered within the monitoring interface 
can stream results directly to the visualization layer as events are identified, 
while annotations generated during exploration are immediately available for downstream 
machine-learning workflows. The following sections describe the system architecture and 
key design decisions that enable this integrated and scalable workflow.

\section{Streaming as a Unifying Design Principle}
\label{sec:principle}
Consider a representative DAS analysis workflow. An analyst loads a directory of 
hourly recordings spanning weeks or months and navigates through continuous data 
streams to identify candidate events. Selected events are annotated directly within 
the streaming viewer, while others are examined in spectral or \emph{f-k} domains 
to refine denoising and processing parameters. Once validated interactively, the 
resulting processing configuration is submitted to a batch executor for large-scale 
analysis of the complete recording. A detection algorithm then scans the processed 
stream, with detected events rendered directly onto the same visualization surface 
used for exploration and annotation.

Each stage of this workflow imposes distinct requirements on the underlying system 
architecture. Interactive browsing demands bounded memory usage independent of 
recording duration. Annotation requires coordinates expressed in stable physical 
units that remain consistent across zoom levels, rendering resolutions, and 
computing environments. Streaming signal processing requires operators to 
preserve internal state—such as IIR filter conditions or running normalization 
statistics—across chunk and file boundaries. Detection algorithms must operate 
continuously over uninterrupted streams and remain robust to events spanning 
adjacent segments or files. Traditional batch-oriented architectures, which 
assume a complete preloaded array as both input and output, cannot satisfy 
these constraints simultaneously.

FiLark addresses these challenges by treating streaming compatibility as a 
first-class architectural principle across every layer of the framework. The 
I/O subsystem presents multi-file acquisitions as a single continuous logical 
stream. The signal-processing backend provides stateful operators with native 
support for boundary-aware and overlap-aware execution, ensuring numerical 
continuity across chunks. The algorithm integration layer defines a standardized 
streaming interface through which detectors consume incremental data and emit 
events in real time. The visualization layer operates directly on the same stream 
abstraction rather than on static preloaded arrays. Because this abstraction is 
shared system-wide, processing pipelines configured interactively on small regions 
of interest can be deployed directly on full-recording batch workflows without 
modification, while annotations and detector outputs remain inherently aligned 
within a common coordinate framework.

\section{Architecture and Implementation}
\label{sec:impl}

\subsection{Overview}
\label{sec:arch}

\begin{figure}[t]
\centering
\includegraphics[width=\linewidth]{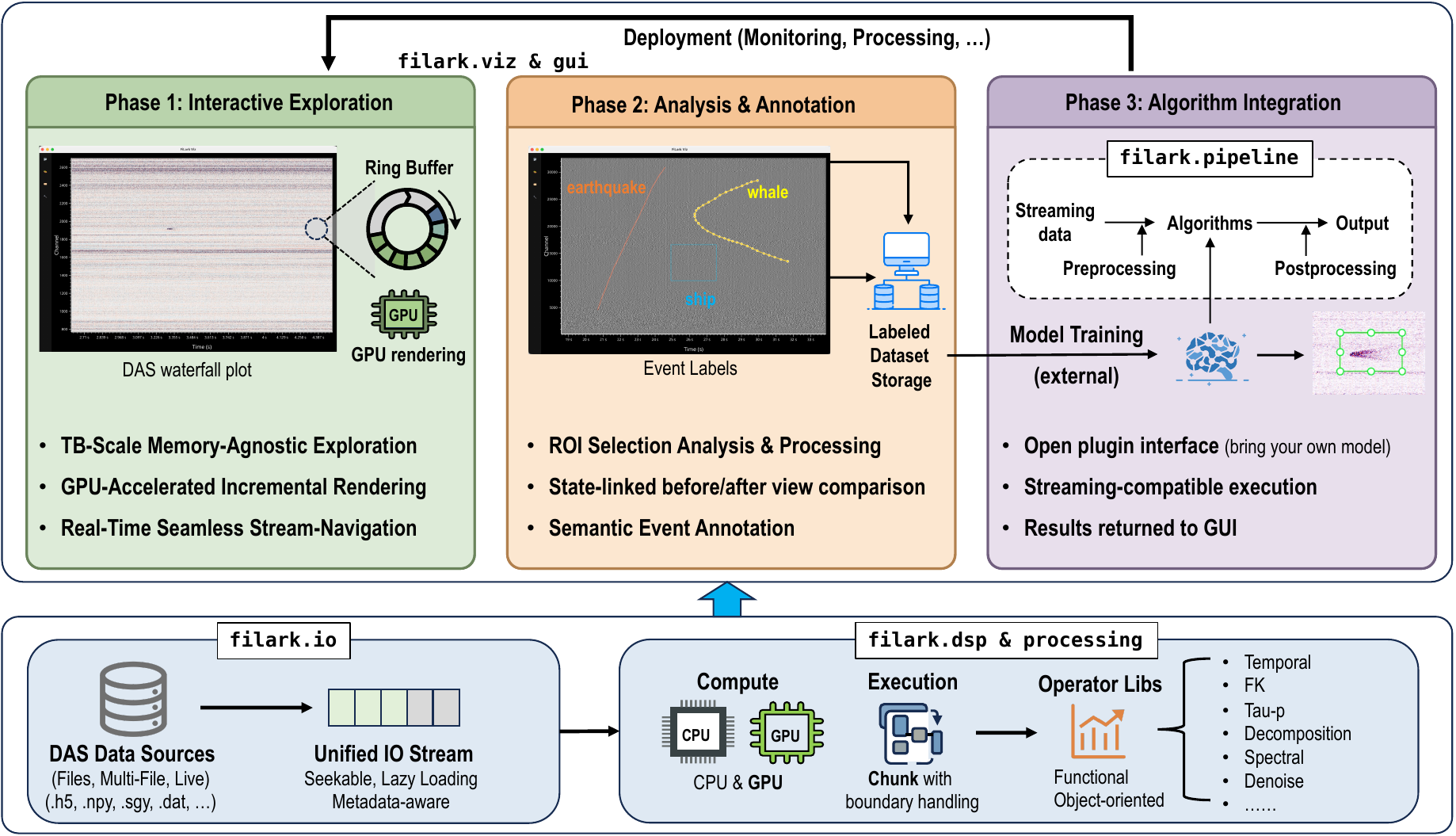}
\caption{Layered architecture of FiLark.
         The framework is organized into three user-facing phases supported
         by two shared backend layers.
         Phase~1 (interactive exploration) uses a ring-buffer GPU renderer
         to browse arbitrarily long records at constant memory.
         Phase~2 (analysis and annotation) provides ROI inspection and
         physical-coordinate event labeling within the same session.
         Phase~3 (algorithm integration) exposes a plugin interface through
         which custom detectors receive streaming data and return results
         to the GUI as canvas overlays.
         The \texttt{filark.io} and \texttt{filark.dsp} layers carry no GUI
         dependency and can be used independently in research scripts or
         headless pipelines.}
\label{fig:architecture}
\end{figure}

FiLark is organized around three user-facing phases supported by shared backend
layers (Fig.~\ref{fig:architecture}).
Phase~1 (interactive exploration) and Phase~2 (analysis and annotation) are 
implemented through the desktop graphical user interface, 
while the remaining core components—including the I/O subsystem, signal-processing 
backend, pipeline executor, and algorithm integration interface—are entirely 
independent of the GUI. These backend layers can therefore be imported directly 
into research scripts, Jupyter notebooks, or headless batch-processing pipelines 
without launching the interactive viewer.
The GUI assembles these components into an interactive application but owns no
numerical algorithms or storage-specific logic.

The I/O and signal-processing layers form the architectural foundation of the 
framework. The I/O subsystem presents all DAS data sources—including continuous 
multi-file acquisitions—through a unified metadata-aware streaming abstraction 
shared consistently across visualization, annotation, processing, and monitoring 
components. Built on top of this abstraction, the signal-processing backend provides 
the operator library, stateful chunked execution engine, and pipeline executor 
required for scalable analysis. The GUI extends these capabilities with the streaming 
renderer, annotation tools, region-of-interest analysis windows, and plugin 
infrastructure, but introduces no independent numerical processing logic. This 
separation is a key architectural feature: the processing pipeline configured 
interactively within the viewer is exactly the same executable object later 
submitted to large-scale batch execution, eliminating the translation and 
reproducibility issues commonly encountered when moving from exploratory analysis 
to production workflows.

\subsection{Ring-buffer streaming visualization}
\label{sec:viz}

\begin{figure}[t]
\centering
\includegraphics[width=\linewidth]{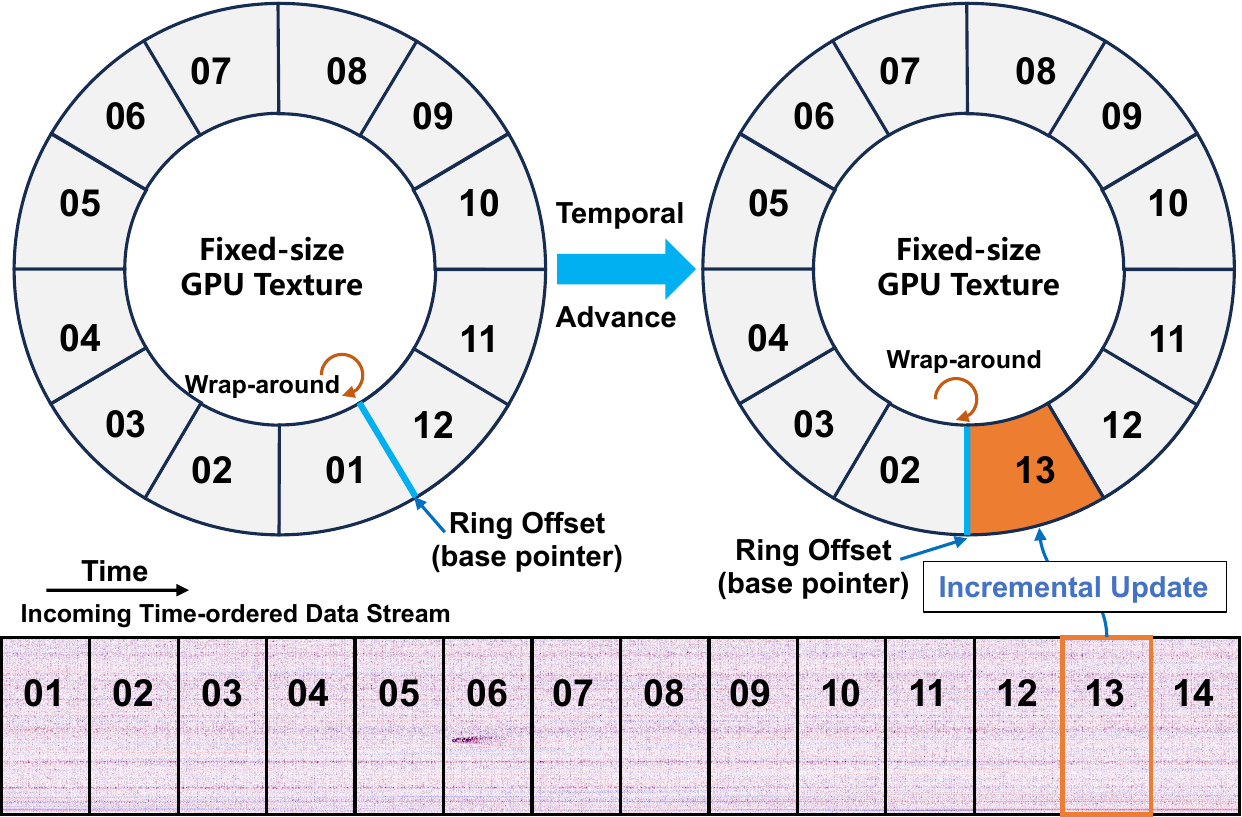}
\caption{Ring-buffer GPU texture mechanism.
         The fixed-size GPU texture is divided into equal-width time slots
         (numbered 01--12 in the left panel).
         When the display advances temporally, only the newly exposed slot
         (slot~13, highlighted in orange) is fetched from the I/O layer and
         written into the vacated region of the ring; all other slots remain
         in GPU memory unchanged.
         A shader-uniform ring offset (blue pointer) tracks the logical start
         of the viewport, so the GPU redraws the same fixed-size texture on
         every frame regardless of how far the record has scrolled.
         Memory footprint is therefore determined by the display resolution
         alone and does not grow with recording duration.}
\label{fig:ringbuffer}
\end{figure}

DAS recordings are long, continuous, and data-intensive, making their interactive
exploration closer to navigating a video stream than opening a static array.
The goal of
FiLark's visualization layer is therefore to provide a video-player-like browsing
experience for DAS archives: analysts can play, seek, zoom, and change channel
ranges without first extracting or preloading fixed-length segments. To support
this experience, the core idea is to keep GPU memory fixed and update only the
newly visible portion of the display. FiLark implements this idea with a
custom OpenGL image renderer built around a fixed-size texture allocated 
once at session initialization. Rather than resizing or reallocating GPU memory 
as navigation proceeds, the texture functions as a ring buffer (Fig.~\ref{fig:ringbuffer}) whose physical 
dimensions correspond only to the active display resolution. The logical viewport 
position within the DAS recording is represented as an offset maintained in 
shader uniforms rather than as a direct pointer into GPU memory. As the user 
navigates along the time axis, the viewport scheduler computes only the newly 
exposed region, retrieves the corresponding data from the I/O layer, and updates 
the vacated segment of the ring buffer. Consequently, the volume of data transferred 
during navigation scales only with the incremental pan distance rather than with 
the total buffer size or recording duration. This design ensures bounded memory 
usage and sustained interactive rendering rates regardless of whether the underlying 
recording spans minutes, days, or weeks. Supplementary Videos~S1--S3
collectively show this constant-memory browsing behavior across exploratory
analysis, large-directory playback, and simulated real-time monitoring.

The scheduler further decouples temporal and spatial navigation to support stable 
interaction during both offline exploration and live-streaming acquisition. 
Time-axis motion is directly coupled to the streaming update mechanism, enabling 
smooth playback and continuous scrolling as new data arrive, while channel-axis 
navigation remains independently controlled through explicit API and keyboard-driven 
commands. This separation is particularly important in real-time monitoring scenarios, 
where the viewer may auto-scroll continuously at acquisition speed while the analyst 
simultaneously adjusts the visible channel range without interrupting the streaming 
state. FiLark therefore prioritizes deterministic keyboard-driven navigation over 
purely mouse-driven interaction, ensuring stable and predictable behavior under 
continuous high-rate data ingestion.

In DAS recordings, weak signals are often masked by strong energy outside
the band of interest. During interactive browsing, analysts therefore need a
temporary spectral-conditioning aid to decide where to inspect, annotate, or
process next.
An interactive display filter is integrated directly into the visualization layer as 
a render-time processing stage applied only to the currently visible buffer. 
Implemented as either a zero-phase or causal IIR filter, this mechanism reuses the signal-processing backend in a streaming
context. The causal IIR mode preserves filter state across successive viewport
updates, avoiding boundary artifacts that would appear if each displayed block
were filtered independently. The filter modifies only the rendered representation 
of the data and leaves the underlying recording unchanged, allowing analysts to 
interactively apply, remove, or adjust spectral conditioning without committing 
to a permanent preprocessing workflow. This capability enables rapid exploration 
of different frequency bands and facilitates visual identification of weak or 
obscured signals during navigation.

The feature is particularly valuable in ocean DAS deployments, where strong 
low-frequency hydrodynamic noise often masks the signal band of interest 
(Fig.~\ref{fig:annotation}).
By enabling real-time spectral conditioning directly within the streaming viewer, 
analysts can dynamically enhance targeted signal components while continuously 
browsing long-duration recordings. Supplementary Videos~S2 and~S3 show the display filter being used during accelerated auto-scroll and simulated real-time monitoring.

\subsection{Annotation and labeled dataset creation}
\label{sec:annot}

\begin{figure}[t]
\centering
\includegraphics[width=\linewidth]{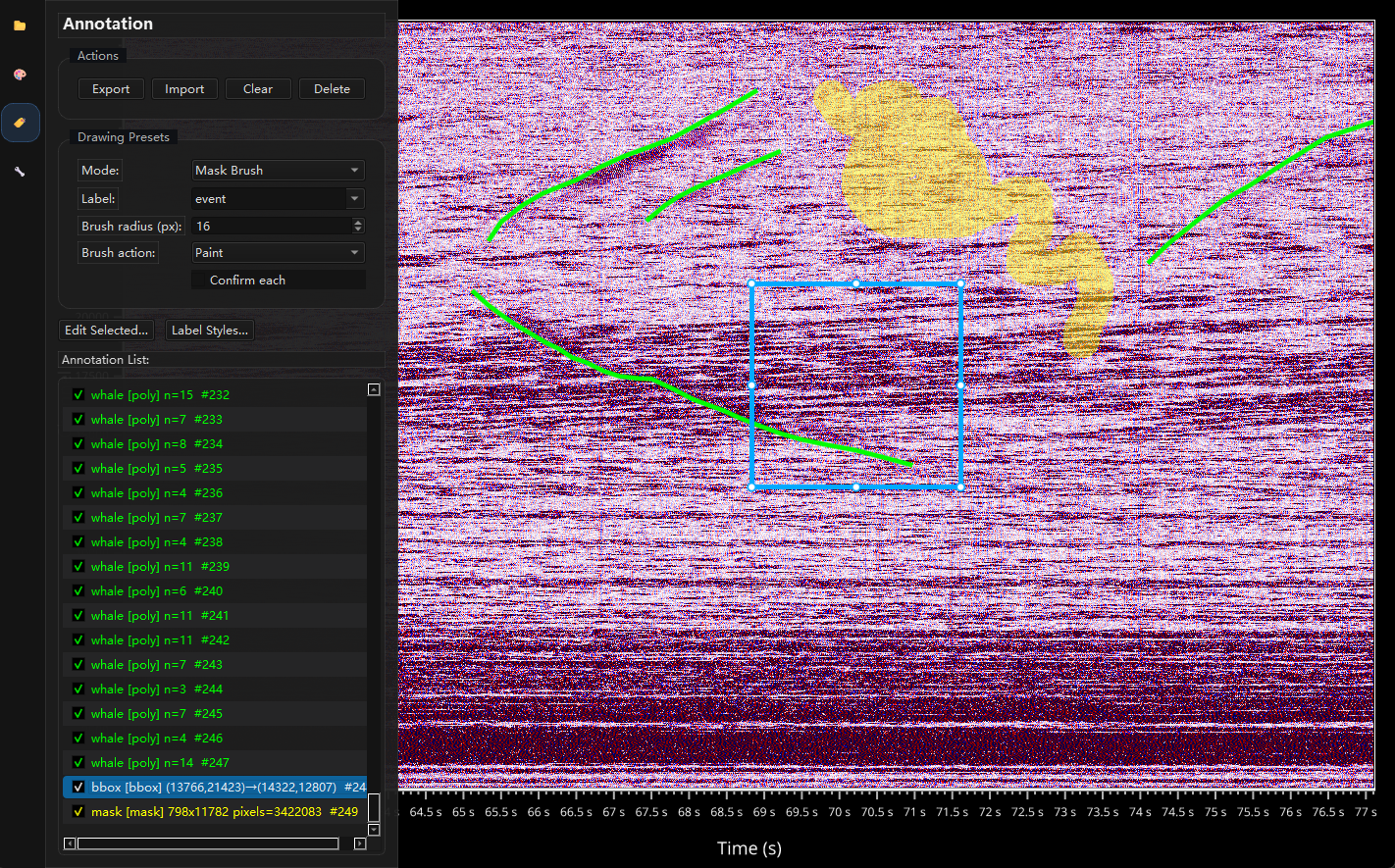}
\caption{The FiLark annotation interface applied to a DAS recording.
	         Bounding boxes, polylines,
	         and masks are drawn directly on the streaming
	         canvas in physical time--channel coordinates.
         The annotation panel (right) lists all labeled items with assigned
         categories, optional confidence scores, and review flags.
         Coordinates are stored in physical units so that annotations remain
         correctly registered when the file is reopened at a different zoom
         level or on a different machine.
         The exported annotation file embeds the source metadata needed to
         extract any labeled data block automatically, without manual
         coordinate conversion.}
\label{fig:annotation}
\end{figure}

DAS acquisitions produce long, spatially dense, and repeatedly observed signal
patterns, making machine-learning methods increasingly attractive to the DAS
community. Realizing this potential, however, depends on the ability to create
large numbers of reliable labels quickly and conveniently from continuous
recordings.
The annotation subsystem sits at the interface between the streaming visualization environment and downstream machine-learning workflows. Its primary goal is to make event labeling within continuous DAS recordings as natural and scalable as the interactive exploration that precedes it. The same interface also supports the replay and manual recheck of monitoring outputs, so algorithm-generated candidates can be corrected and folded back into subsequent model iteration. Because the viewer itself supports navigation across recordings of arbitrary duration, the annotation system is designed to operate seamlessly at the same scale.

Annotations are created interactively within the streaming canvas using three geometric primitives: axis-aligned bounding boxes, free-form polylines, and masks(Fig.~\ref{fig:annotation}).
Bounding boxes are well suited for compact localized events, such as transient acoustic disturbances or discrete seismic arrivals. Polylines, by contrast, are designed for signals whose signatures evolve continuously across the time–channel domain, including seismic moveout trajectories and Doppler-like patterns generated by moving vessels, which cannot be represented accurately by rectangular regions alone. Masks support spatially extended or irregular event regions and are also useful for reviewing algorithm-generated detection overlays. All annotation coordinates are stored in physical units—time in seconds and position in channel index—rather than in display-space pixels. This choice is fundamental to reproducibility and correctness rather than merely a convenience. An annotation generated at one zoom level or display resolution must map to exactly the same physical region when reopened on another machine or under a different visualization configuration, a guarantee that pixel-referenced annotations cannot provide.

Each annotation object stores a categorical label together with optional metadata including confidence scores, review status, and free-text notes. Annotation sessions are exported as plain-text records that embed a complete source description, including file paths, acquisition dimensions, sample rate, channel spacing, and UTC start time. This metadata allows FiLark to automatically reconstruct the original data source and extract the precise data region associated with any annotation, optionally with user-defined temporal and spatial padding. As a result, labeled event extraction requires no manual coordinate conversion or format-specific scripting and produces consistently normalized output arrays independent of original storage format, dimensional ordering, or downsampling configuration.

The practical outcome is a direct and reproducible path from exploratory browsing of continuous DAS recordings to the generation of machine-learning-ready labeled datasets. The same interactive session in which an analyst discovers candidate events, annotates them, and verifies label quality simultaneously produces the structured data required for training and evaluation of detection or classification models. Supplementary Video~S2 demonstrates bbox and polyline labeling, annotation export, and later import for review; Supplementary Video~S3 additionally demonstrates monitoring results imported for manual recheck.

\subsection{Signal processing, pipelines, and algorithm integration}
\label{sec:dsp}

FiLark is more than a visualization tool.
The signal processing, pipeline, and algorithm layers carry no GUI dependency
and constitute a standalone DAS research library that can be imported directly
in scripts, Jupyter notebooks, or headless processing workflows without
launching the desktop application.
A practical motivation for this independence is scale: at several terabytes per
day, a DAS recording cannot in general be loaded into memory before an algorithm
is applied.
Chunked, overlap-aware processing is therefore a structural requirement inherent
to DAS research---not an optional optimization---and the framework enforces it
by design rather than delegating it to the user.

The signal-processing backend implements the principal operator families commonly 
used in DAS analysis and is designed specifically for continuous, large-scale 
recordings. Its architecture follows two key principles that enable robust 
streaming execution. First, operators whose behavior depends on temporal 
context—such as IIR filters and running normalization methods—are implemented as 
\emph{stateful-compatible} operators that preserve internal state across successive calls. 
This allows arbitrarily long recordings to be processed incrementally in 
sequential chunks without introducing discontinuities at chunk boundaries. 
The IIR filtering suite further includes causal, sample-wise real-time variants 
(\texttt{iir\_filter\_time\_rt} together with corresponding bandpass, lowpass, and highpass 
wrappers) intended for true streaming deployment. These operators accept incremental 
data chunks and return outputs of matching length while maintaining correct state 
propagation internally, enabling the same implementation to operate directly on 
live acquisition streams without algorithmic modification.
Second, spatially coupled operators—including local singular value decomposition (SVD) denoising, \emph{f-k}  fan 
filtering, and directional transforms—support overlap-aware chunked execution. 
In this model, each processing block is temporarily expanded with additional 
neighboring channels, or halo regions, before computation and trimmed back 
afterward prior to output generation. This strategy preserves numerical continuity 
and suppresses boundary artifacts that would otherwise emerge at segment edges 
during large-scale streaming execution.

The operator library covers the principal transform families commonly encountered in DAS
analysis (Table~\ref{tab:dsp}).
All operators are available as functional CPU implementations that accept and
return standard numerical arrays, making them suitable for direct use in research
scripts and Jupyter notebooks without any GUI dependency.
A class-based transform interface wraps these functions in composable objects
that can be assembled into reusable processing chains and passed either to the
streaming viewer's preprocessing hooks or to the pipeline executor.
GPU-accelerated implementations are available for computationally intensive
operators, including fast Fourier transform (FFT)-based filtering, \emph{f-k} transforms, truncated
SVD, and cross-correlation, through a PyTorch backend \citep{paszke2019pytorch}.
The same processing chain can switch between CPU and GPU execution without any
configuration changes.
Because CPU and GPU implementations share the same operator interface,
acceleration can be enabled without changing the processing graph or the
chunk-boundary handling.
As shown in Table~\ref{tab:comparison}, GPU-accelerated signal processing is
not provided by the open-source DAS frameworks compared here; on filter-intensive
workloads it can reduce end-to-end runtimes by one to two orders of magnitude
(Section~\ref{sec:perf:dsp}).

\begin{table}[t]
\caption{Signal processing operator categories in FiLark.
         All operators are available as stateless functional CPU implementations.
         \emph{Stateful}: a causal streaming variant is implemented that carries
         filter state (e.g.\ IIR initial conditions) across chunk boundaries.
         \emph{GPU}: a GPU-accelerated implementation is available via PyTorch.}
\label{tab:dsp}
\centering
\small
\begin{tabular}{p{0.20\linewidth} p{0.44\linewidth} p{0.12\linewidth} p{0.10\linewidth}}
\toprule
Category & Operations & Stateful & GPU \\
\midrule
Temporal filtering
  & IIR (offline \& realtime second-order sections, SOS), finite impulse response (FIR)(offline \& realtime),
    FFT-domain bandpass/stop,
    running automatic gain control (AGC), running mean, Gaussian smooth, median smooth
    (all with stateful streaming variants),
    de-mean, de-trend, z-score,
    spectral whitening, one-bit normalisation, amplitude clipping
  & \checkmark & \checkmark \\
Spatial
  & Channel resampling, spatial gradient / Laplacian,
    running mean, Gaussian smooth, gain equalisation
    (streaming exponential moving average, EMA variant), de-mean
  & \checkmark & \checkmark \\
\emph{f-k} domain
  & Fan (velocity) filter, directional wedge denoising
  & --- & \checkmark \\
Decomposition
  & Local-window SVD denoising, truncated SVD reconstruction
  & --- & \checkmark \\
Radon / $\tau$-$p$
  & Linear and parabolic Radon transforms,
    sparse inversion-based denoising
  & --- & \checkmark \\
Cross-correlation
  & Gather, all-pairs, phase cross-correlation,
    ambient-noise cross-correlation pipeline
  & --- & \checkmark \\
Nonlinear
  & Anisotropic diffusion / Perona--Malik structure-tensor filter
  & --- & \checkmark \\
Spectral analysis
  & Short-time Fourier transform (streaming chunk-boundary variant),
    Welch power spectral density (PSD), root-mean-square (RMS) envelope, spectral centroid
  & \checkmark & \checkmark \\
\bottomrule
\end{tabular}
\end{table}

The pipeline layer wraps any signal processing chain in a unified execution
interface that handles chunking logic, halo management, and output serialization.
A pipeline object is a standalone Python component with no GUI dependency:
it accepts a Tape source and a sequence of operators, applies them in a
chunk-wise loop with correct overlap and state propagation, and writes the
result to disk or returns it in memory.
The same operator chain used in a script or Jupyter notebook can therefore
be submitted to a full-length out-of-core run without any code changes---the
pipeline hides the complexity of chunking, file stitching, and boundary
management analogously to the pipeline abstraction in machine-learning
frameworks such as Hugging Face Transformers \citep{wolf2020transformers}.

The framework exposes several extension points for user-defined components.
The \texttt{Tape} protocol defines the minimal interface a data source must
implement---\texttt{shape}, \texttt{dtype}, \texttt{fs}, \texttt{dx}, and
array-style indexing---so that custom loaders for proprietary interrogator
formats integrate transparently with the viewer, pipeline, and annotation
system.
The monitor interface specifies the contract between a streaming detection
algorithm and the GUI overlay system: any callable that consumes a data
window and returns a list of time--channel-bounded events qualifies as a
monitor, whether it is a threshold-based energy detector, a matched filter,
or a trained neural network.
The GUI processor plugin interface allows new signal processing methods to
be registered with their own parameter specification and exposed through the
existing ROI panel without modifying the core application.

The processing stack extends naturally to automated detection and continuous
monitoring through a \emph{monitor} interface.
A monitor is a streaming-compatible algorithm that consumes a DAS source and
emits detected events---each described by a time and channel extent and a
semantic label---delivered as canvas overlays in real time.
The interface makes no assumption about the detection strategy: threshold-based
energy detectors, matched filters, and trained neural network models are all
valid implementations, provided the algorithm returns events in the expected
format.
Because monitor output shares the same physical coordinate system as manual
annotations, detected events can be reviewed and accepted by an analyst and
promoted directly into the annotation store.
This closes the feedback loop shown in Fig.~\ref{fig:architecture}: labeled data
combining manual annotations and accepted detections can train an improved model
registered as a new monitor---completing the cycle from raw continuous recording
to a data-driven DAS monitoring system within a single environment.

\subsection{Streaming-aware I/O layer}
\label{sec:io}

All three core components described above---the streaming viewer, the annotation
system, and the processing library---depend on a common foundation: a
data access layer that presents any DAS source, regardless of format, storage
layout, or whether the acquisition spans one file or several hundred, as a
continuous, metadata-rich stream.
The central interface provides uniform access to the physical dimensions of a
DAS source: sample rate, channel spacing, dimension ordering, and optional UTC
acquisition start time.
Concrete implementations cover Hierarchical Data Format version 5 (HDF5) files produced by common DAS interrogators,
NumPy arrays from intermediate processing steps, and raw binary recordings with
externally specified geometry.

A key design decision is the treatment of multi-file acquisitions.
In operational deployments, the interrogator typically writes a new file every
few minutes or hours, and a complete acquisition is a folder of such files.
FiLark stitches compatible files into a single logical time axis, treating
per-file length variability transparently so that the rest of the stack sees a
continuous stream rather than a sequence of disjoint arrays.
This is essential for stateful operators, which must propagate internal state
across file boundaries, and for event detectors, which must not miss events
that straddle two consecutive files.

At the visualization level, the I/O layer serves view-dependent windows: only
the samples falling within the current viewport are fetched from storage, and
logical downsampling is applied to match the display resolution.
Multiple downsampling strategies accommodate different signal characteristics:
drop mode preserves transient peaks, peak mode maintains amplitude envelopes,
and linear interpolation smooths slowly varying wavefields.
This view-dependent serving is what allows the viewer's memory footprint to
remain independent of dataset size. In practice, panning or zooming therefore
requests only a new physical slice of the source rather than forcing the archive
to be opened as a full array, which keeps browsing responsive even for long
multi-file recordings.

\begin{lstlisting}[caption={Custom binary loader implemented as a FiLark plugin.
  The class declares a \texttt{suffixes} tuple and a \texttt{load} method that
  returns a \texttt{BinTape}.
  \texttt{BinTape} maps the file via \texttt{numpy.memmap}: no data are loaded
  into RAM upfront; only the samples that fall within the current viewport are
  read on each render tick.
  Passing the loader to \texttt{GuiConfig} is sufficient;
  FiLark applies it to each matching file in a folder and stitches the results
  into a continuous logical stream automatically.},
  label=lst:binloader, float=t]
import numpy as np
from pathlib import Path
from filark.io import BinTape
from filark.gui import GuiConfig, run_app

class BinLoader:
    # Filename: DAS-{gauge}-{fs}-{dx}-{ch0}-{nc}-{timestamp}.dat
    suffixes = ('.bin', '.dat')

    def load(self, path):
        p = Path(path)
        parts = p.stem.split('-')
        fs, dx, nc = float(parts[2]), float(parts[3]), int(parts[5])
        nt = p.stat().st_size // (nc * 4)   # float32 samples
        # BinTape wraps the file as np.memmap -- zero copy, lazy access
        return BinTape(str(p), nt=nt, nc=nc,
                       dtype=np.float32, dims='nt_nc',
                       fs=fs, dx=dx)

# Register the loader; FiLark handles folder stitching automatically
cfg = GuiConfig(loaders=[BinLoader()])
run_app(cfg=cfg)
\end{lstlisting}

\subsection{Region-of-interest analysis}
\label{sec:roi}

\begin{figure}[t]
\centering
\includegraphics[width=\linewidth]{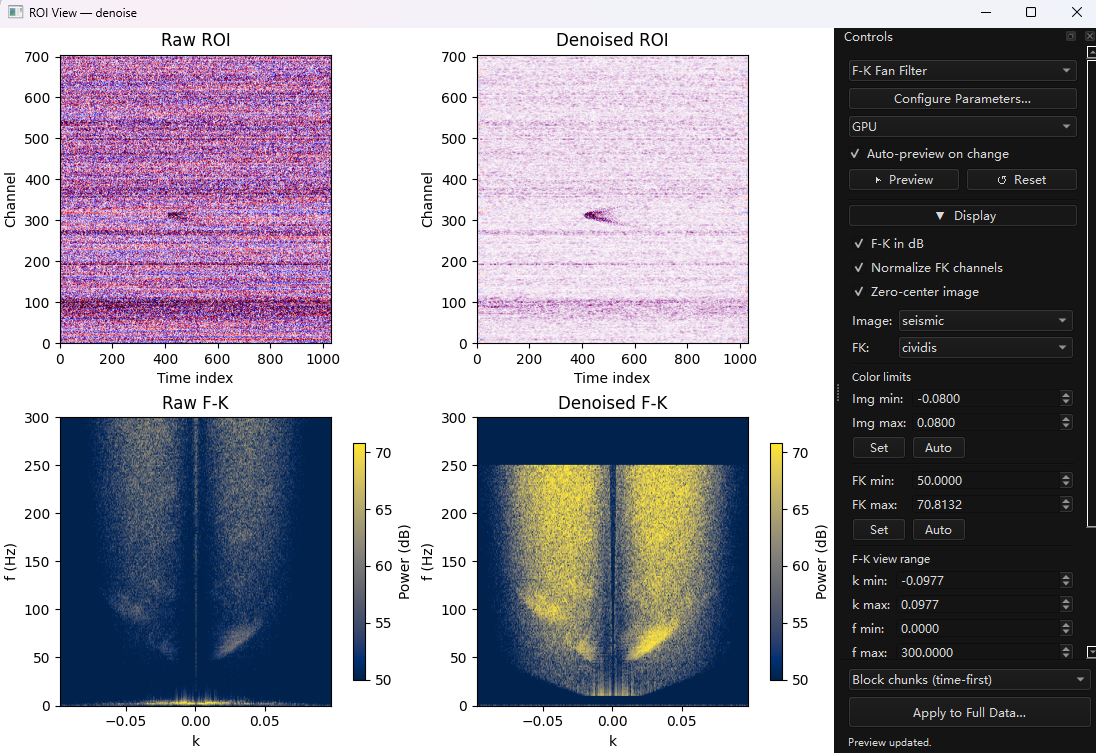}
\caption{Region-of-interest (ROI) analysis window in FiLark.
         A segment selected on the streaming canvas is immediately available
         for spectral, structural, and processing analysis without leaving the
         viewing session.
         The ROI window provides an FK view of the selected segment,
         a denoising preview with raw and processed panels side-by-side,
         and method and parameter controls.
         Processing configurations tuned on a small ROI can be escalated to
         full-source chunked execution through the same interface,
         with no code duplication.}
\label{fig:roi}
\end{figure}

Interactive DAS interpretation frequently requires local quantitative analysis
of a candidate event before a processing or labeling decision can be made.
The practical challenge is that these local checks are often needed while
the analyst is still browsing a much larger continuous record, where exporting
segments to separate scripts breaks context and slows iteration. FiLark therefore
turns region-of-interest (ROI) selection into an in-session analysis bridge. The same selected event
can be inspected, processed locally, and, when needed, escalated to full-source
execution without leaving the viewer.
FiLark supports this through ROI windows that open directly
from a selection drawn on the main canvas (Fig.~\ref{fig:roi}).
Rather than requiring the analyst to export a segment and process it in a
separate environment, the ROI window provides immediate access to spectral and
structural analysis within the same session.

Available ROI analyses include \emph{f-k} spectrum inspection, short-time Fourier
transform, power spectral density (PSD), channel-to-channel cross-correlation,
$\tau$-$p$ (Radon) transform, SVD-based signal and noise separation, and
root-mean-square (RMS)
trace analysis.
The denoising ROI view is particularly useful for processing parameter tuning:
the raw and processed versions of the selected segment are displayed side by
side, so the effect of a processing choice can be judged on the event itself.
When full-record processing is needed, the same configuration can be submitted
to the pipeline executor for out-of-core processing of the full source. This
tight coupling between local
inspection and full-data execution is a direct expression of the streaming-first
design. The configuration tested on a small window and the configuration
deployed on the complete recording are identical objects, made possible by the
shared operator interface across the interactive and batch layers. Supplementary Video~S1 demonstrates this
ROI-centered loop, including \emph{f-k} inspection, local filtering, full-data
application, and linked comparison between the original and processed sources.

\subsection{Extensibility and plugin model}
\label{sec:ext}

DAS workflows are rarely standardized across deployments. Interrogator-specific
file formats, project-specific preprocessing conventions, and domain-specific
detection algorithms often need to coexist within the same exploration
environment. FiLark is therefore designed as an extensible framework rather than
a closed application, allowing project-specific loaders, processors, monitors,
and analysis views to be added without modifying the core viewer.

Custom file loaders can be registered with the application to support
non-standard formats or interrogator-specific naming conventions without
modifying the I/O layer.
A loader implements two attributes: a \texttt{suffixes} tuple that identifies
the file types it handles, and a \texttt{load} method that returns a
\texttt{Tape} object given a file path.
When the user opens a folder, FiLark applies the registered loader to each
matching file and stitches the results into a continuous logical stream.
Listing~\ref{lst:binloader} shows a complete loader for a proprietary binary
DAS format whose acquisition parameters are encoded in the filename.
The loader returns a \texttt{BinTape}, which wraps the file as a
\texttt{numpy.memmap}: data are never loaded into RAM upfront, and only the
samples needed for the current viewport are fetched on each render tick.
The implementation is under 20~lines, and no changes to the core viewer or
I/O layer are required.

Three plugin categories extend the analytical capabilities of the GUI.
\emph{Processor plugins} expose custom preprocessing methods in the ROI panel.
Each plugin declares its parameters in a schema, and the interface renders
the corresponding controls automatically.
\emph{Monitor plugins} implement the streaming detector interface
(Section~\ref{sec:dsp}) and appear in the detection menu alongside built-in
algorithms.
\emph{Analysis plugins} attach custom figures or widgets to the current
data source, providing an extension point for domain-specific outputs such as
travel-time curves, beamforming images, or custom spectrograms.

Stream preprocessing and postprocessing hooks allow transform chains to be
applied to every block served to the main viewer.
These hooks accept metadata-aware factory functions, so parameters that
depend on sample rate or channel spacing are recomputed automatically when
the source changes.
A project-specific FiLark deployment can integrate custom formats, algorithms,
and visualizations without modifying the core viewer, and the resulting
extensions are compatible with any standard FiLark installation.

\section{End-to-End Workflow}
\label{sec:workflow}

To make the design concrete, we trace a representative browsing and annotation
session on ocean-cable DAS data.
Ocean DAS cables simultaneously record seismic signals, ship traffic, marine
biological sounds, and strong low-frequency hydrodynamic noise
\citep{sladen2019distributed, lior2021detection}.
Without spectral conditioning, dominant noise renders most signals invisible,
and a deployed submarine cable can accumulate weeks to months of continuous
recordings.

A folder of hourly recordings is opened as a single continuous logical stream.
A render-time bandpass display filter is engaged to suppress the dominant
low-frequency noise; the filter can be adjusted or toggled at any point
without modifying the stored data.
The analyst navigates the full record using keyboard-driven pan and zoom,
scanning for candidate events at interactive frame rates despite the
multi-terabyte source.

When a cluster of candidate arrivals is identified, bounding boxes are drawn
directly on the canvas and assigned categorical labels.
For arrivals with coherent moveout across channels, polylines capture the
signal trajectory.
An ROI window is opened on a representative segment to inspect the spectral
and \emph{f-k} character of the event.
The same spectral conditioning applied interactively in the ROI window can
be assembled into a pipeline configuration and submitted as an offline batch
job over the full source.

Supplementary Video~S3 demonstrates this workflow in a simulated real-time
monitoring setting. Rather than integrating a specific detector, the demo uses a
monitor plugin to emulate an algorithm with controlled latency, taking
approximately 900~ms to process each 1~s data block with additional random
jitter before returning randomly generated bbox, polyline, and mask events.
These results are delivered through FiLark's asynchronous monitor API, so the
viewer continues to scroll and the display filter remains responsive while
algorithm-like results are written to a result file and overlaid on the canvas.
The saved result file is then reopened through the annotation module for manual
recheck.

\section{Relationship to Existing DAS Software}
\label{sec:comparison}

Table~\ref{tab:comparison} compares FiLark with DASPy, DASCore, and xdas, three
open-source Python frameworks developed for DAS data analysis.
All three provide useful foundations for data handling and signal processing,
including multi-file access and batch array processing.
The comparison therefore focuses on the additional capabilities required for an
interactive continuous-record workflow: constant-memory browsing, render-time
display filtering, annotation on the streaming canvas, monitor feedback, causal
state propagation, and GPU-accelerated signal processing.
Xpick, the companion phase-picking tool in the xdas ecosystem, is marked as
partial for annotation because it operates on pre-selected bounded segments
rather than on a continuously navigated record.

\begin{table}[t]
\caption{Capability comparison of open-source DAS Python frameworks considered
         in this work, including
         DASPy~\citep{hu2024daspy}, DASCore~\citep{chambers2024dascore},
         and xdas~\citep{trabattoni2025xdas}.
         A dash indicates that the capability was not identified in the public
         paper, documentation, or repository pages reviewed.
         $^{a}$xdas includes Xpick, a companion web application for phase
         picking on pre-selected bounded segments;
         continuous-record navigation is not supported.}
\label{tab:comparison}
\centering
\small
\begin{tabular}{lcccc}
\toprule
Capability & FiLark & DASPy & DASCore & xdas \\
\midrule
Multi-file streaming I/O
		                          & \checkmark & \checkmark & \checkmark & \checkmark \\
Constant-memory interactive viewer
                          & \checkmark & ---        & ---        & ---        \\
Interactive display filter (render-time, zero-latency)
                          & \checkmark & ---        & ---        & ---        \\
Interactive event annotation
                          & \checkmark & ---        & ---        & partial$^{a}$ \\
Algorithm integration with GUI feedback
                          & \checkmark & ---        & ---        & ---        \\
GPU signal-processing operators
                          & \checkmark & ---        & ---        & ---        \\
Causal state across chunks
					                          & \checkmark & \checkmark & ---        & \checkmark    \\
Signal processing operators
                          & \checkmark & \checkmark & \checkmark & \checkmark \\
Batch array processing    & \checkmark & \checkmark & \checkmark & \checkmark \\
\bottomrule
\end{tabular}
\end{table}

\section{Performance Evaluation}
\label{sec:perf}

\subsection{Streaming renderer}
\label{sec:perf:render}

To characterize real-time throughput, we benchmarked the streaming renderer
in the most demanding operating mode, \emph{fit-all} scrolling, where the display
spans all channels simultaneously with the IIR display filter active.
Benchmarks were run on a MacBook Air with an Apple M2 processor (8~GB unified
memory), using HDF5 files as the data source.
The renderer buffer was fixed at $2048 \times 4096$ pixels and scrolled at
$1\times$ real-time speed over 10~s of data.
All timing components are expressed in ms per second of DAS data (ms~s$^{-1}$);
a total below 1000~ms~s$^{-1}$ indicates sustained real-time throughput.

\begin{figure}[t]
\centering
\includegraphics[width=\linewidth]{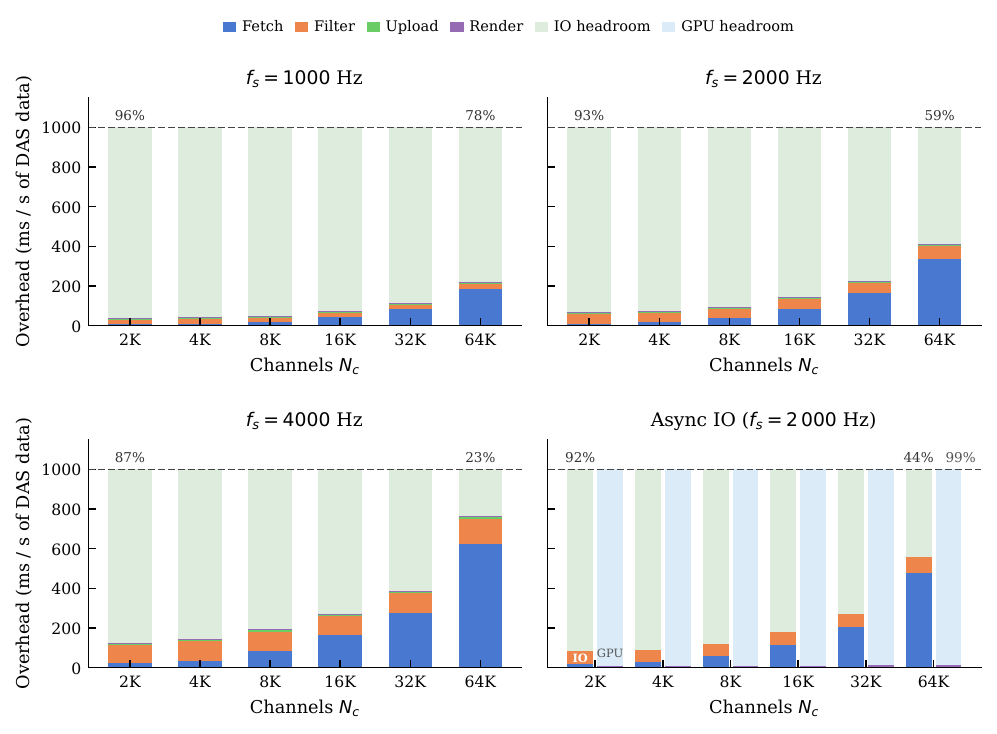}
\caption{Streaming renderer throughput as a function of channel count $N_c$
         and sampling rate $f_s$ (synchronous mode, top three panels).
         Stacked bars show the normalized per-component overhead in
         ms per second of DAS data; the dashed line marks the 1000~ms
         real-time budget.
         Percentages indicate remaining headroom at the smallest and
         largest $N_c$.
         The bottom-right panel shows the measured async-IO breakdown at
         $f_s = 2000$~Hz: left bars represent the IO thread
         (fetch~$+$~filter); right bars represent the GPU thread
         (upload~$+$~render), which runs in parallel with the IO thread
         and therefore retains ${\approx}99\%$ of the 1-s budget as
         headroom regardless of $N_c$.
         Test platform: Apple M2 MacBook Air (8~GB unified memory),
         HDF5 source, display filter enabled,
         renderer buffer $2048 \times 2048$ pixels.}
\label{fig:perf}
\end{figure}

Three structural observations emerge from Fig.~\ref{fig:perf}.
First, the Render cost is approximately 5~ms/s and remains constant across
all $(n_c, f_s)$ combinations.
The GPU always redraws the same fixed-size texture regardless of the underlying
data dimensions, confirming that the ring-buffer design isolates draw-call
cost from both channel count and sample rate.
Second, the Filter cost depends only on $f_s$ (approximately 24~ms/s at
1\,000~Hz, 47~ms/s at 2\,000~Hz, and 97~ms/s at 4\,000~Hz) and is
independent of $n_c$.
The IIR filter operates on the 2048-pixel render buffer rather than on the
full channel array, so physical channel count does not affect filter latency.
Third, Fetch is the primary scaling bottleneck, growing with $n_c \times f_s$
because fit-all mode must read all channel samples within the visible window.
This is an inherent cost of displaying all channels simultaneously and is not
a deficiency of the I/O layer.

A fourth observation concerns the effect of asynchronous IO on algorithm
headroom.
In synchronous mode, all four components execute serially on a single thread,
so the time budget available for downstream detection algorithms is
\begin{equation}
  H_{\mathrm{sync}} = 1000 - (t_{\mathrm{fetch}} + t_{\mathrm{filter}}
                              + t_{\mathrm{upload}} + t_{\mathrm{render}})
                      \;\text{ms\,s}^{-1}.
\end{equation}
In asynchronous mode, the IO thread (fetch~$+$~filter) and the GPU thread
(upload~$+$~render) run concurrently on separate threads.
The effective latency is $\max(t_{\mathrm{IO}},\, t_{\mathrm{GPU}})$, and a
detection algorithm dispatched on a third thread can overlap its execution with
both.
As long as each thread individually remains within budget---i.e.\
$t_{\mathrm{IO}} < 1000$~ms\,s$^{-1}$ and $t_{\mathrm{GPU}} < 1000$~ms\,s$^{-1}$---the
algorithm inherits the \emph{full} 1-s wall-clock window:
\begin{equation}
  H_{\mathrm{async}} = 1000 \;\text{ms\,s}^{-1}.
\end{equation}
The bottom-right panel of Fig.~\ref{fig:perf} illustrates this: the GPU thread
occupies only ${\approx}11$~ms\,s$^{-1}$ regardless of $N_c$, leaving
${\approx}99\%$ headroom on that thread alone.
Even at $N_c = 65{,}536$ the IO thread retains 44\% headroom, so both
conditions hold and a co-running algorithm receives the full budget.

These results confirm that the streaming renderer sustains real-time
throughput across a wide range of DAS acquisition configurations on a laptop
without a discrete GPU.
In practice, interactive exploration rarely requires continuous full-speed
scrolling: the analyst typically pauses, steps, and zooms rather than
watching the record scroll at acquisition speed.
The benchmark therefore represents a stress test of the rendering engine,
not a typical operating condition.
The practical implication is that frame rate remains high and consistent
throughout a browsing session, so that annotation---which requires accurate
cursor placement on individual arrivals---is not impeded by rendering latency.

\subsection{Signal processing performance}
\label{sec:perf:dsp}

\begin{figure}[t]
\centering
\includegraphics[width=\linewidth]{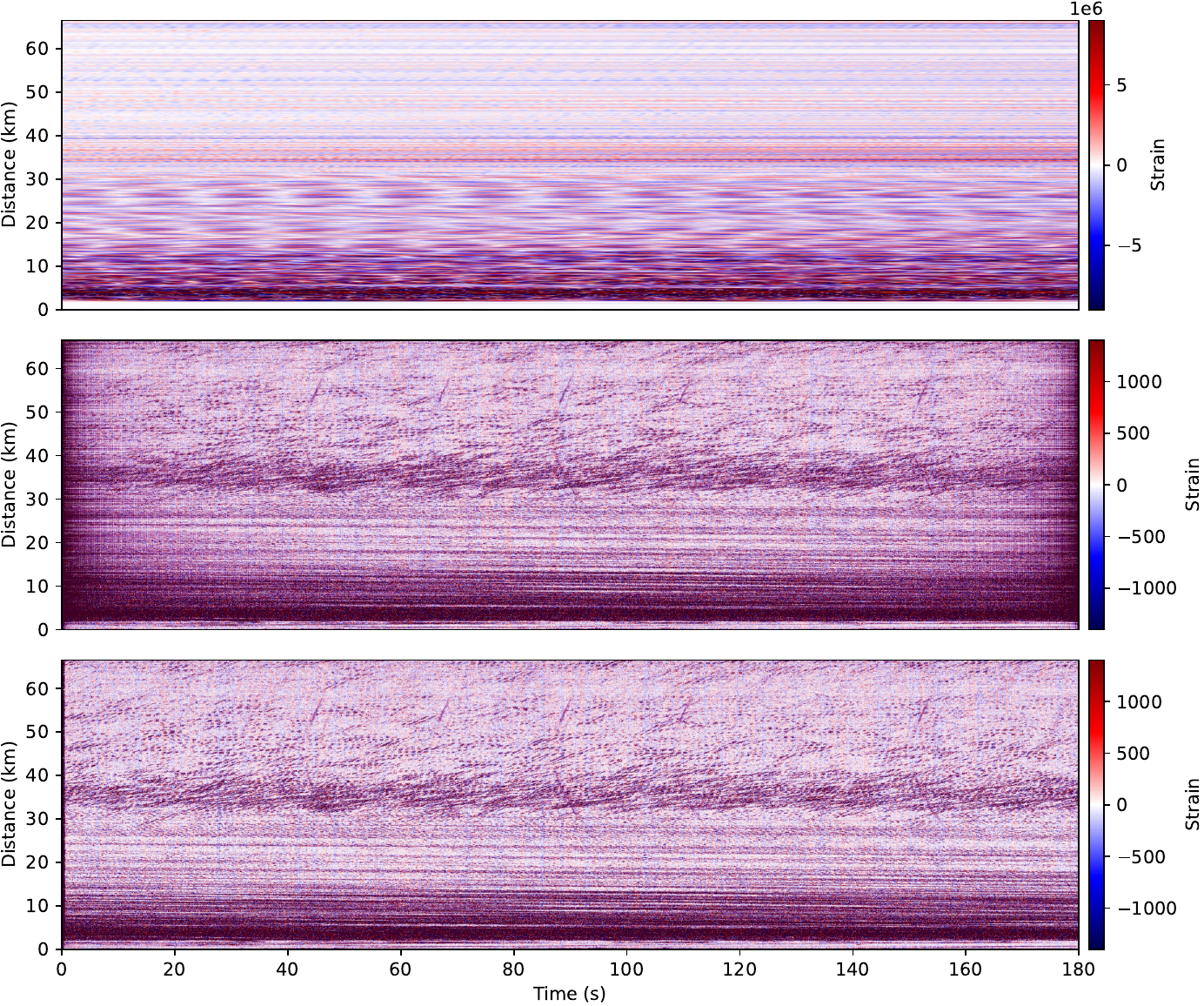}
\caption{OOI ocean-cable DAS data \citep{wilcock2023distributed} processed with FiLark's GPU-accelerated
         signal processing backend.
         (a)~Raw DAS data showing strong low-frequency hydrodynamic noise.
         (b)~Full FFT bandpass filtering (15--32~Hz), which suppresses the
         noise floor and reveals seismic and acoustic arrivals.
         (c)~Causal stateful IIR bandpass filtering (15--32~Hz) applied
         chunk-by-chunk with filter state preserved across boundaries,
         matching the streaming execution model.
         All three panels show the same time--channel window on the same
         colour scale.
         Input data are three concatenated OOI files accessed as a single
         logical Tape.}
\label{fig:ooi}
\end{figure}

To evaluate the GPU signal processing backend on real DAS data, we benchmarked
FFT-based and causal IIR bandpass filtering on three concatenated Ocean Observatories Initiative (OOI) ocean-cable
files processed as a single logical stream via the Tape abstraction
\citep{wilcock2023distributed}.
Fig.~\ref{fig:ooi} illustrates the effect of the two filter modes on the raw data.
End-to-end GPU acceleration reduced runtime from 25.7~s to 6.7~s (3.8$\times$)
for full FFT-based filtering and from 44.3~s to 3.4~s (13$\times$) for
overlap-add chunked filtering; pure filter-kernel acceleration reached
50.8$\times$ and 392$\times$, respectively.

The causal stateful IIR filter addresses a different use case: real-time
interactive visualization.
Because it preserves filter state across chunk boundaries, it can be applied
to an incrementally arriving stream and produce a filtered output without
boundary discontinuities.
In practice, the first chunk processed after the filter is initialized
carries a short transient determined by the filter order and the initial
condition---subsequent chunks are free of this artifact because state is
carried forward exactly.
Its sustained throughput of 97.1~Msamples/s is sufficient for real-time
rendering at all channel counts tested in Section~\ref{sec:perf:render},
confirming that stateful IIR filtering can serve as the render-time display
filter without imposing any additional latency on the browsing session.

\section{Conclusions}
\label{sec:conc}

DAS recordings present a class of software challenges that are not well addressed
by existing batch-oriented DAS toolboxes.
The fundamental source of these challenges is architectural: when the
software abstraction does not match the continuous, streaming character of the
data, every stage of the workflow---browsing, preprocessing, detection,
and labeling---requires bespoke solutions that cannot easily be composed or
reused.
FiLark addresses this by establishing streaming compatibility as a first-class
design contract across the entire software stack, from file access to GPU
rendering.

The consequences of this design are not limited to visualization performance.
A streaming-aware I/O layer that stitches multi-file acquisitions into a
continuous logical stream, combined with signal processing operators that
propagate state across chunk boundaries, means that the same configuration
used for interactive inspection is directly deployable for production batch
processing.
An annotation system embedded in the streaming viewer---storing events in
physical rather than pixel coordinates and exporting source metadata alongside
labels---removes the principal barrier to creating labeled DAS datasets at
practical scale.
An algorithm integration layer that expresses DAS detectors as streaming
monitors, and connects their output to the same annotation infrastructure,
establishes a feedback loop from raw data to labeled training sets to improved
detectors that operates entirely within the framework.

The broader contribution of FiLark is therefore not a collection of new DAS
algorithms, but a software architecture in which the streaming nature of DAS
data is treated as the organizing principle rather than as a special case.
This architecture is designed to remain productive as the scale of DAS
deployments grows and as data-driven detection methods become more central
to DAS research.

\section*{Code availability}

FiLark is released under the MIT License, with source code and documentation
available at \url{https://github.com/JintaoLee-Roger/filark} and
\url{https://filark.readthedocs.io}.

\section*{Declaration of competing interest}

The authors declare no known competing financial interests or personal
relationships that could have appeared to influence the work reported in
this paper.

\printcredits

\bibliographystyle{cas-model2-names}
\bibliography{refs}

\end{document}